\documentclass[twocolumn,aps,superscriptaddress]{revtex4-1}
\usepackage[latin9]{inputenc}
\usepackage{color}
\usepackage{amsmath}
\usepackage{amssymb}
\usepackage{graphicx}
\usepackage[unicode=true,pdfusetitle,
 bookmarks=true,bookmarksnumbered=false,bookmarksopen=false,
 breaklinks=false,pdfborder={0 0 0},backref=false,colorlinks=true]
 {hyperref}
\hypersetup{
 colorlinks,linkcolor=red,citecolor=blue}
\usepackage{breakurl}

\makeatletter

\@ifundefined{textcolor}{}
{%
 \definecolor{BLACK}{gray}{0}
 \definecolor{WHITE}{gray}{1}
 \definecolor{RED}{rgb}{1,0,0}
 \definecolor{GREEN}{rgb}{0,1,0}
 \definecolor{BLUE}{rgb}{0,0,1}
 \definecolor{CYAN}{cmyk}{1,0,0,0}
 \definecolor{MAGENTA}{cmyk}{0,1,0,0}
 \definecolor{YELLOW}{cmyk}{0,0,1,0}
}

\usepackage{amsfonts}

\makeatother

\begin{document}

\title{Strongly-coupled nanotube electromechanical resonators}

\author{Guang-Wei Deng}

\email{These authors contributed equally to this work.}

\affiliation{Key Laboratory of Quantum Information, University of Science and
Technology of China, Chinese Academy of Sciences, Hefei 230026, China}

\affiliation{Synergetic Innovation Center of Quantum Information \& Quantum Physics,
University of Science and Technology of China, Hefei, Anhui 230026,
China}

\author{Dong Zhu}

\email{These authors contributed equally to this work.}

\affiliation{Key Laboratory of Quantum Information, University of Science and
Technology of China, Chinese Academy of Sciences, Hefei 230026, China}

\affiliation{Synergetic Innovation Center of Quantum Information \& Quantum Physics,
University of Science and Technology of China, Hefei, Anhui 230026,
China}

\author{Xin-He Wang}

\email{These authors contributed equally to this work.}

\affiliation{State Key Laboratory of Low-Dimensional Quantum Physics, Department
of Physics \& Tsinghua-Foxconn Nanotechnology Research Center, Tsinghua
University, Beijing 100084, China}

\affiliation{Collaborative Innovation Center of Quantum Matter, Beijing 100084,
China}

\author{Chang-Ling Zou}

\email{These authors contributed equally to this work.}

\affiliation{Key Laboratory of Quantum Information, University of Science and
Technology of China, Chinese Academy of Sciences, Hefei 230026, China}

\affiliation{Synergetic Innovation Center of Quantum Information \& Quantum Physics,
University of Science and Technology of China, Hefei, Anhui 230026,
China}

\author{Jiang-Tao Wang}

\affiliation{State Key Laboratory of Low-Dimensional Quantum Physics, Department
of Physics \& Tsinghua-Foxconn Nanotechnology Research Center, Tsinghua
University, Beijing 100084, China}

\affiliation{Collaborative Innovation Center of Quantum Matter, Beijing 100084,
China}

\author{Hai-Ou Li}

\affiliation{Key Laboratory of Quantum Information, University of Science and
Technology of China, Chinese Academy of Sciences, Hefei 230026, China}

\affiliation{Synergetic Innovation Center of Quantum Information \& Quantum Physics,
University of Science and Technology of China, Hefei, Anhui 230026,
China}

\author{Gang Cao}

\affiliation{Key Laboratory of Quantum Information, University of Science and
Technology of China, Chinese Academy of Sciences, Hefei 230026, China}

\affiliation{Synergetic Innovation Center of Quantum Information \& Quantum Physics,
University of Science and Technology of China, Hefei, Anhui 230026,
China}

\author{Di Liu}

\affiliation{Key Laboratory of Quantum Information, University of Science and
Technology of China, Chinese Academy of Sciences, Hefei 230026, China}

\affiliation{Synergetic Innovation Center of Quantum Information \& Quantum Physics,
University of Science and Technology of China, Hefei, Anhui 230026,
China}

\author{Yan Li}

\affiliation{Key Laboratory of Quantum Information, University of Science and
Technology of China, Chinese Academy of Sciences, Hefei 230026, China}

\affiliation{Synergetic Innovation Center of Quantum Information \& Quantum Physics,
University of Science and Technology of China, Hefei, Anhui 230026,
China}

\author{Ming Xiao}

\affiliation{Key Laboratory of Quantum Information, University of Science and
Technology of China, Chinese Academy of Sciences, Hefei 230026, China}

\affiliation{Synergetic Innovation Center of Quantum Information \& Quantum Physics,
University of Science and Technology of China, Hefei, Anhui 230026,
China}

\author{Guang-Can Guo}

\affiliation{Key Laboratory of Quantum Information, University of Science and
Technology of China, Chinese Academy of Sciences, Hefei 230026, China}

\affiliation{Synergetic Innovation Center of Quantum Information \& Quantum Physics,
University of Science and Technology of China, Hefei, Anhui 230026,
China}

\author{Kai-Li Jiang}

\affiliation{State Key Laboratory of Low-Dimensional Quantum Physics, Department
of Physics \& Tsinghua-Foxconn Nanotechnology Research Center, Tsinghua
University, Beijing 100084, China}

\affiliation{Collaborative Innovation Center of Quantum Matter, Beijing 100084,
China}

\author{Xing-Can Dai}

\affiliation{State Key Laboratory of Low-Dimensional Quantum Physics, Department
of Physics \& Tsinghua-Foxconn Nanotechnology Research Center, Tsinghua
University, Beijing 100084, China}

\affiliation{Collaborative Innovation Center of Quantum Matter, Beijing 100084,
China}

\author{Guo-Ping Guo}

\email{Corresponding author: gpguo@ustc.edu.cn}

\affiliation{Key Laboratory of Quantum Information, University of Science and
Technology of China, Chinese Academy of Sciences, Hefei 230026, China}

\affiliation{Synergetic Innovation Center of Quantum Information \& Quantum Physics,
University of Science and Technology of China, Hefei, Anhui 230026,
China}

\maketitle
\textbf{Coupling an electromechanical resonator with carbon-nanotube
quantum dots is a significant method to control both the electronic
charge and the spin quantum states. By exploiting a novel micro-transfer
technique, we fabricate two strongly-coupled and electrically-tunable
mechanical resonators on a single carbon nanotube for the first time.
The frequency of the two resonators can be individually tuned by the
bottom gates, and strong coupling is observed between the charge states
and phonon modes of each resonator. Furthermore, the conductance of
either resonator can be nonlocally modulated by the phonon modes in
the other resonator. Strong coupling is observed between the phonon
modes of the two resonators, which provides an effective long distance
electron-electron interaction. The generation of phonon-mediated-spin
entanglement is also analyzed for the two resonators. This strongly-coupled
nanotube electromechanical resonator array provides an experimental
platform for studying the coherent electron-phonon interaction, the
phonon mediated long-distance electron interaction, and entanglement
state generation.}

\section{Introduction}

Carbon nanotubes (CNTs) \cite{Saito1998} are noted for their nearly
perfect structures with nanometer diameter, ultralow mass density,
great mechanical strength and elastic properties, as well as ballistic
electron transport \cite{Ilani2010,Laird2015}. Owing to good electrical
conductivity and lack of impurities and net nuclear spin, the electron
charge and spin states in gate-defined CNT quantum dots (QDs) \cite{Biercuk2005,Sapmaz2006a,Sapmaz2006,Grove-Rasmussen2008,Jung2013}
are promising candidates for solid-state quantum information processing.
However, a scalable quantum processor requires long-range couplings,
which is a challenge for QDs, because there are only local interactions
between neighboring QDs. Many researches have been undertaken on the
development of a ``quantum bus'' to transfer quantum information,
carried by electrons, over certain distances \cite{Petersson2012,Deng2015}.
For example, a single electron can be conveyed between QDs over distances
of micrometers \cite{Meunier2011,Mcneil2011}, and an integrated superconducting
microwave cavity can mediate the coupling between spins over distances
of millimeters \cite{Petersson2012}.

On the other hand, the excellent mechanical properties of CNTs enable
their use as high frequency and high-quality-factor nanomechanical
resonators \cite{Laird2012,Moser2014}. The vibrations of suspended
CNTs can modulate the electrochemical potential of quantum dots, which
leads to coherent coupling between single electron charge and phonon
\cite{Steele2009,Lassagne2009}. Additionally, the deformation of
CNTs can induce an effective transverse magnetic field applied on
the electron spins that arises from the spin-orbit interaction \cite{Ando2000,Huertas2006,Kuemmeth2008},
thereby allowing spin flips by phonons \cite{Palyi2012,Ohm2012}.
These approaches provide avenues toward the coherent operation and
transduction of the quantum state of CNT QDs by a phonon, or alternatively,
the electronic manipulation of the phonon quantum state \cite{Wang2014,Bulaev2008,Rudner2010,Benyamini2014,Zippilli2009,Meerwaldt2012}.
Theoretically, mechanically-induced two-qubit gates and maximally-entangled
states for two spins trapped in a single CNT have recently been studied
\cite{Wang2015}. However, those previous works have only focused
on the localized electron-phonon interactions. Hence, the great potential
of using phonons as flying qubits for communicating electron spins
over long-distance \cite{Rabl2010,Gustafsson2014,Schuetz2015} is
overlooked.

Here we demonstrate a highly-tunable electrically-coupled nanomechanical
resonator system of a single CNT with two suspended sections. We developed
a novel transfer method, which can precisely posit the CNT to the
designated location and maintain the clean surface of the CNT without
requiring of chemical treatment. In the conductance spectrum, an avoided
crossing indicates a strong coupling between the two CNT electromechanical
resonators and the hybridization of two modes, and also proves the
strong coupling between the electron charge and individual hybrid
mode. To our knowledge, this is the first demonstration of non-local
coupling between an electron charge and phonon in a carbon nanotube.
Our theoretical study also predicts that remote-entanglement spin-state
preparation is feasible in this device, for practical experimental
parameters.

\begin{figure}[tp]
\centerline{\includegraphics[width=1\columnwidth]{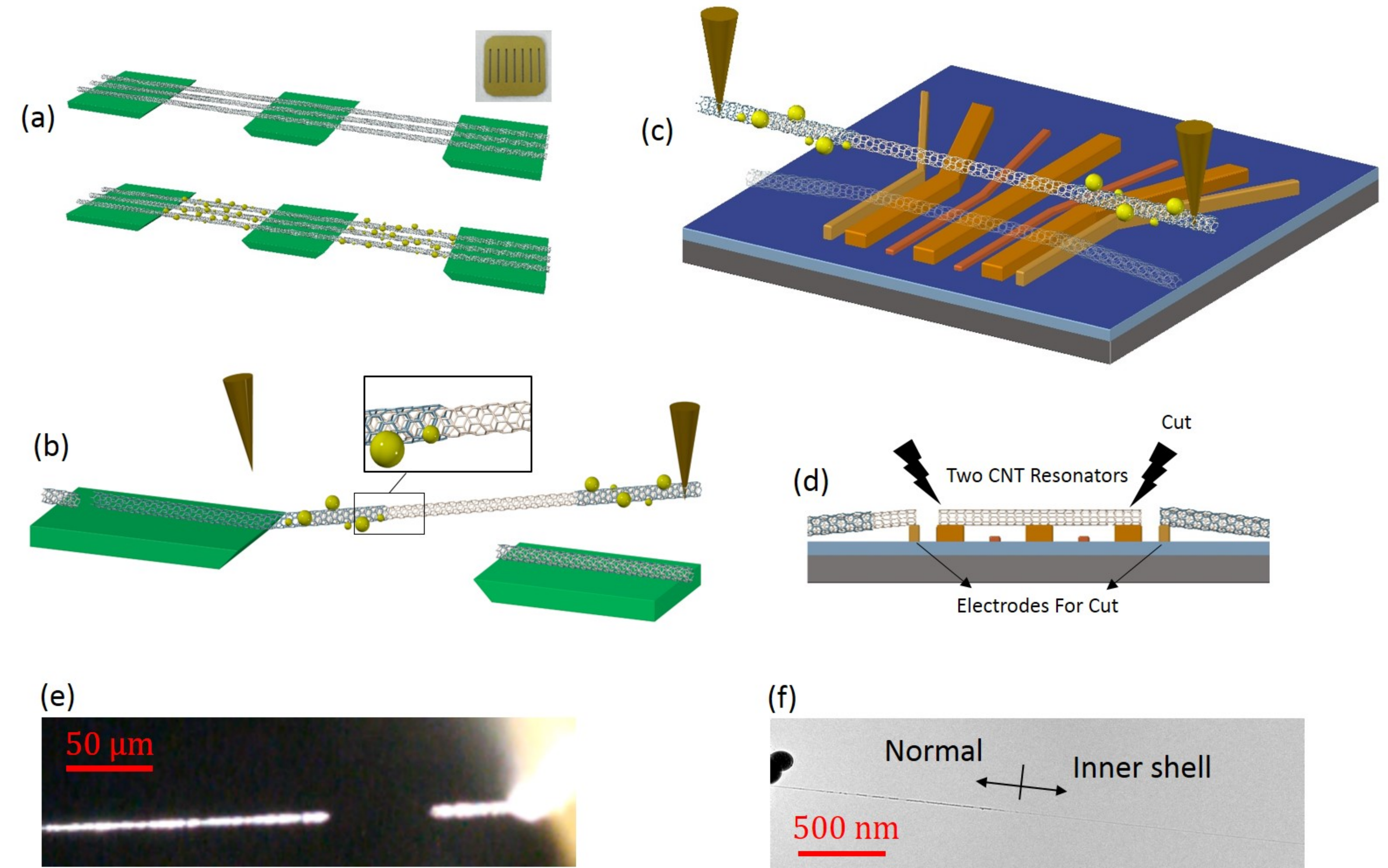}} \protect\protect\protect\caption{\textbf{Transfer method} (a) Growth of the CNT using the ethane CVD
method. The long, parallel CNTs are grown on a silicon substrate and
suspended over trenches. Typically, the CNTs are double- or triple-
walled. To obtain a direct visualization, we then deposit some TiO$_{2}$
nano-particles on the suspended parts of the CNTs. (b) With an optical
microscope and two homemade tips, we cut and take off the outer shell
of the suspended part of the CNT, maintaining the inner part, which
is ultra clean. (c) After straining the individual CNT between the
two tips, we place it onto the designed metal contacts, which have
been biased with about 3-5 V for attracting the CNT. (d) Locally cutting
off the redundant parts of the CNT. (e) Micrograph of the CNT, corresponding
to Fig. 1(b), where the bright parts contain TiO$_{2}$ and the outer
shell. The invisible part between the two bright parts only contains
the inner shell. (f) Transmission electron micrograph of the CNT.
Normal part corresponds to the bright parts in Fig. 1(e) and the inner
shell corresponds to the invisible part.}
\end{figure}

\section{Results}

\subsection{Experiment Setup}

Figure 1 shows the sample fabrication method, where a CNT (typically
single- or double-walled, 2-3 nm in diameter and grown by chemical
vapour deposition) is suspended over two trenches (1.2 $\mu$m wide,
200 nm deep) between three metal (Ti/Au) electrodes. The CNT is transferred
by a novel near-field micro-manipulation method, by which the perfect
clean single CNTs are deterministically and precisely posited on the
electrodes, without degrading the quality of the CNT (see Fig. 1).
The measurements are performed in a He3 refrigerator at a base temperature
of approximately 270 mK and at pressures below $10^{-6}$ torr (see
Methods for details).

The suspended CNT is biased and actuated by two electrodes underneath
the CNT (Fig. 2(a)). Each suspended section of the CNT simultaneously
serves as both a mechanical resonator and a quantum dot. The gate
voltages, $V_{g1(2)}$, induce an average additional charge $\left\langle q_{i}\right\rangle =C_{gi}V_{gi}\,(i=1,2)$
on the CNT, where $C_{gi}$ is the capacitance between the $i$-th
gate and the CNT. The attraction between the charge $q_{i}$ and its
opposite charge $-q_{i}$ on the $i$-th gate causes an electrostatic
force downward on the CNT, leading to a mean electrostatic force on
the CNT as 
\begin{equation}
F_{i}=\frac{\partial\left\langle U_{i}\right\rangle }{\partial z_{i}}=\frac{1}{2}\frac{\partial C_{gi}}{\partial z}(V_{gi}^{{\rm DC}}+\delta V_{gi})^{2}.
\end{equation}
Here, $\frac{\partial C_{gi}}{\partial z}$ is the derivative of the
gate capacitance with respect to the distance between the gates and
the CNT, while $V_{gi}^{{\rm DC}}$ and $\delta V_{gi}$ are the DC
bias and AC signal electric fields applied to the electrodes, respectively.

By applying a DC voltage $V_{gi}^{{\rm DC}}$, the nanobeam-type nanomechanical
resonator can be deformed by the static force $F_{i}^{{\rm DC}}=\frac{1}{2}\frac{\partial C_{gi}}{\partial z}(V_{gi}^{{\rm DC}})^{2}$,
and the induced additional tension on the CNT changes the frequencies
of the mechanical resonances. In addition, the electron transport
properties of the quantum dot also depend on the electrochemical potential
on the dot, and so are controllable by the DC voltage. For instance,
Figs. 2(b, c) show the currents $I_{1(2)}$ through the quantum dot
as a function of gate voltage. The $1$st quantum dot is working in
the Coulomb blockade regime, while the $2$nd is working in the Fabry-Perot
interference regime \cite{Liang2001}. Both quantum dots can be tuned
to work in different regimes by changing the $V_{gi}^{{\rm DC}}$
\cite{Grove-Rasmussen2007,Moser2014}. If an RF driving field $\delta V_{gi}(t)=\delta V_{gi}^{{\rm RF}}\cos(2\pi f_{gi}t)$
is applied, when the frequency $f_{gi}$ approaches the resonance
frequency $f_{0i}=\omega_{m,i}/2\pi$ of the $i$-th resonator, the
periodic driving force $F_{i}^{{\rm AC}}=\frac{\partial C_{gi}}{\partial z}V_{gi}^{{\rm DC}}\delta V_{gi}(t)$
will effectively actuate the mechanical vibration. The phonons can
also be generated by a parametric driving force $F_{i}^{{\rm para}}=\frac{1}{2}\frac{\partial C_{gi}}{\partial z}(\delta V_{gi}^{{\rm RF}}){}^{2}\cos^{2}(2\pi f_{gi}t)$
with $f_{gi}=f_{0i}/2$.

\subsection{Individual nanotube resonators}

Before studying the coupled resonators, we first investigate the two
mechanical resonators independently. Owing to the RF driving force,
we obtain the driving displacement vibration $\delta z(\omega,t)=A(\omega)\cos(\omega t+\phi)$,
with driving frequency $\omega$, amplitude of the mechanical oscillator
\begin{align}
A(\omega)= & \frac{\frac{\partial C_{gi}}{\partial z}V_{gi}^{{\rm DC}}\delta V_{gi}^{{\rm RF}}}{m_{{\rm eff}}}\times\nonumber \\
 & \frac{1}{\sqrt{(\omega_{m,i}^{2}-\omega^{2}+\frac{3}{4}\frac{\alpha}{m_{{\rm eff}}}A(\omega)^{2})^{2}+\frac{\omega_{m,i}^{2}\omega^{2}}{Q_{i}^{2}}}},
\end{align}
phase factor $\phi$, effective mass $m_{{\rm eff}}$ and nonlinear
Duffing term $\alpha$. The displacement-modulated capacitor of the
suspended CNT can modify the current, which has the same effect on
the modulated gate voltage as $V_{{\rm eff,i}}(\omega,t)=\frac{V_{gi}^{{\rm DC}}}{C_{gi}}\frac{\partial C_{gi}}{\partial z}\delta z(\omega,t)$
\cite{Huttel2009,Moser2014}. Therefore, the drain-source current
changes with time as $I_{{\rm SD,i}}(t)=\sum_{n}\frac{1}{n!}\frac{d^{n}I_{{\rm SD,i}}^{{\rm DC}}(V_{g,i})}{dV_{g,i}^{n}}[V_{{\rm eff,i}}(\omega,t)]^{n}$,
with the $I_{{\rm SD,i}}^{{\rm DC}}$ shown in Fig. 2. The measured
change of the DC current is approximated by 
\begin{equation}
\Delta I_{sd,i}\approx\frac{1}{4}\frac{d^{2}I_{{\rm SD,i}}^{{\rm DC}}(V_{g,i})}{dV_{g,i}^{2}}[\frac{V_{g,i}^{{\rm DC}}}{C_{g,i}}\frac{\partial C_{g,i}}{\partial z}A(\omega)]^{2}
\end{equation}
to second order. Figure 2 shows the measured DC current as a function
of driving frequency at low temperature, for various driving powers.
For relatively low driving powers, the spectra show symmetry peaks.
The Duffing coefficient $\alpha$ is measured to be of the order of
$10^{12}kg/(m^{2}s^{2})$. High order nonlinearities begin to exist
when the driving power is larger than -40 dBm. (see the supplementary
materials).

\begin{figure}[tp]
\centerline{\includegraphics[width=1\columnwidth]{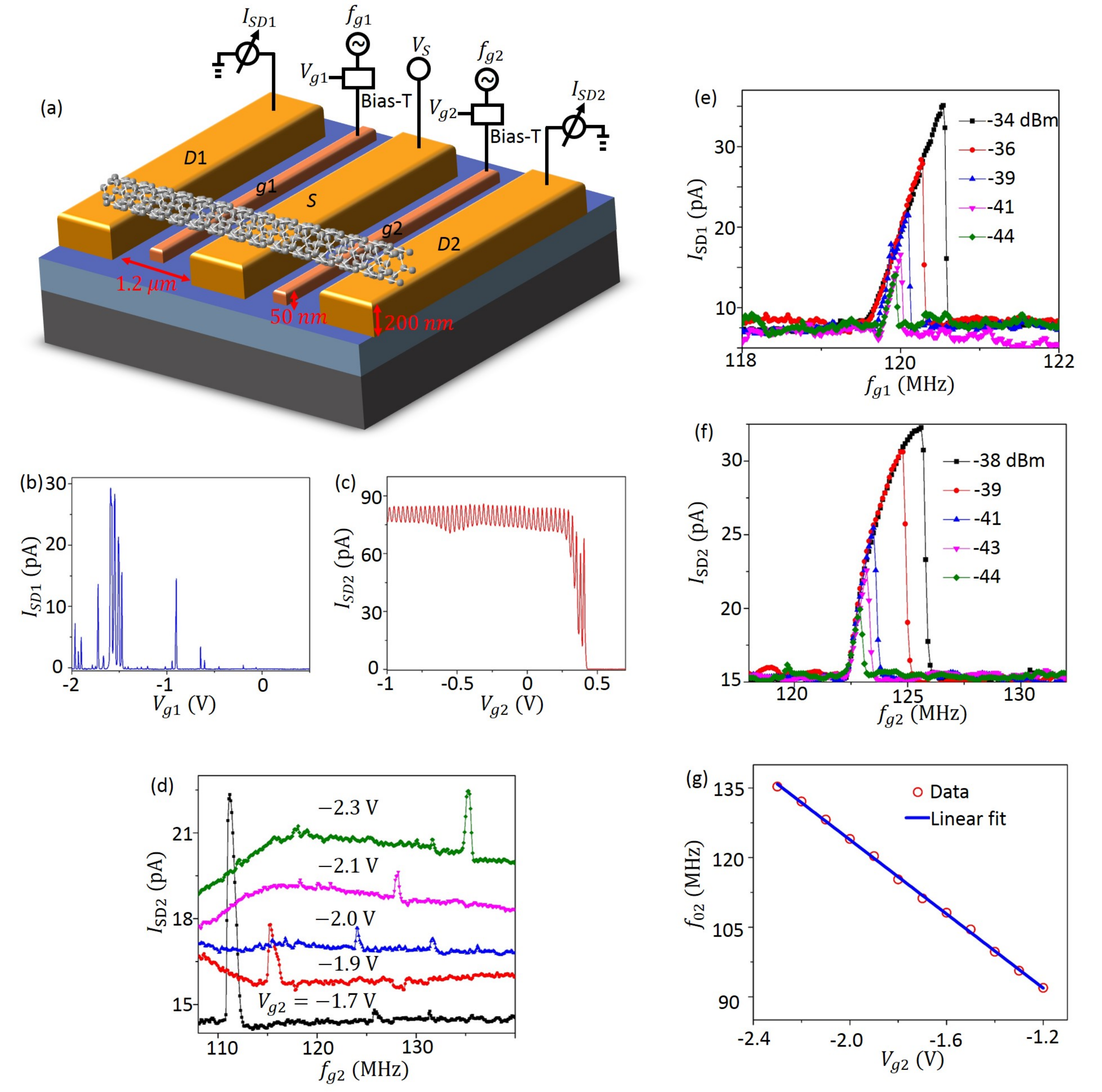}} \protect\protect\protect\caption{\textbf{Measurement circuit and the mechanical vibrations modifying
the current in the quantum dots.} (a) Schematic diagram of the coupled
nanomechanical resonator and double quantum dot system. A single-walled
carbon nanotube is transferred onto three 200 nm height Ti/Au electrodes,
working as sources/drains. Thus, the three source/drain structure
forms two coupled resonators in series. Two 50 nm Ti/Au electrodes
are used as back gates to apply DC and AC voltages. (b-c) Electron
transport properties of the two quantum dots. Both quantum dots are
Coulomb blockaded at large positive gate voltages. Resonator 1 works
like a quantum dot in the large-voltage range while resonator 2 works
in the Fabry-Perot interference regime. Both resonators are biased
at 5 mV.(d) Resonance current peaks of resonator 2 as a function of
its driving frequency, for various gate voltages (noted in the figure).
(e, f) The current spectrum versus driving frequency for various driving
powers. (g) Resonance frequency of resonator 2 as a function of its
gate voltage. The blue line is a linear fit, with an R-squared value
of 99.9\%. }
\end{figure}

The quality factor $Q$ of the resonator and resonance frequency of
the CNT resonator are determined by fitting the spectrum obtained
at low driving power with a Lorentzian function. The quality factors
of both resonators are $\sim10^{4}$ (largest value in our measurements),
which yield an energy relaxation time of $13\thinspace\mathrm{\mu s}$.
Figs. 2(d, g) show the broadband tunability of the resonance frequency
$f_{02}=\omega_{m,2}/2\pi$, which is linearly increased with $V_{g2}^{{\rm DC}}$.
This can be explained as an incremental increase of the elastic tension
on the nanotube, which is almost linearly proportional to the perturbative
DC voltage applied to the gate. From the data for resonator 2, we
fit the coefficient $df_{02}/{dV_{g2}^{{\rm DC}}}=40\thinspace\mathrm{MHz/V}$.
This number is orders of magnitude larger than those reported for
other systems, such as 2 kHz/V for tunability with capacitive forces
\cite{Rugar1991}, 240 kHz/V with a Lorentz force \cite{Karabalin2009},
40 kHz/V for piezoelectric NEMSs \cite{Mahboob2008}, and 10 kHz/V
for a dielectric force setup \cite{Unterreithmeier}. Such a large
frequency-shifting coefficient, as demonstrated by our CNT nanomechanical
device, allows us to tune the phonon modes to be on-resonance or off-resonance
with each other, offering a great ability to reconfigure the phonon-electron
system to regions inaccessible in other systems.

\subsection{Coupled resonators}

\begin{figure}[tp]
\centerline{\includegraphics[width=1\columnwidth]{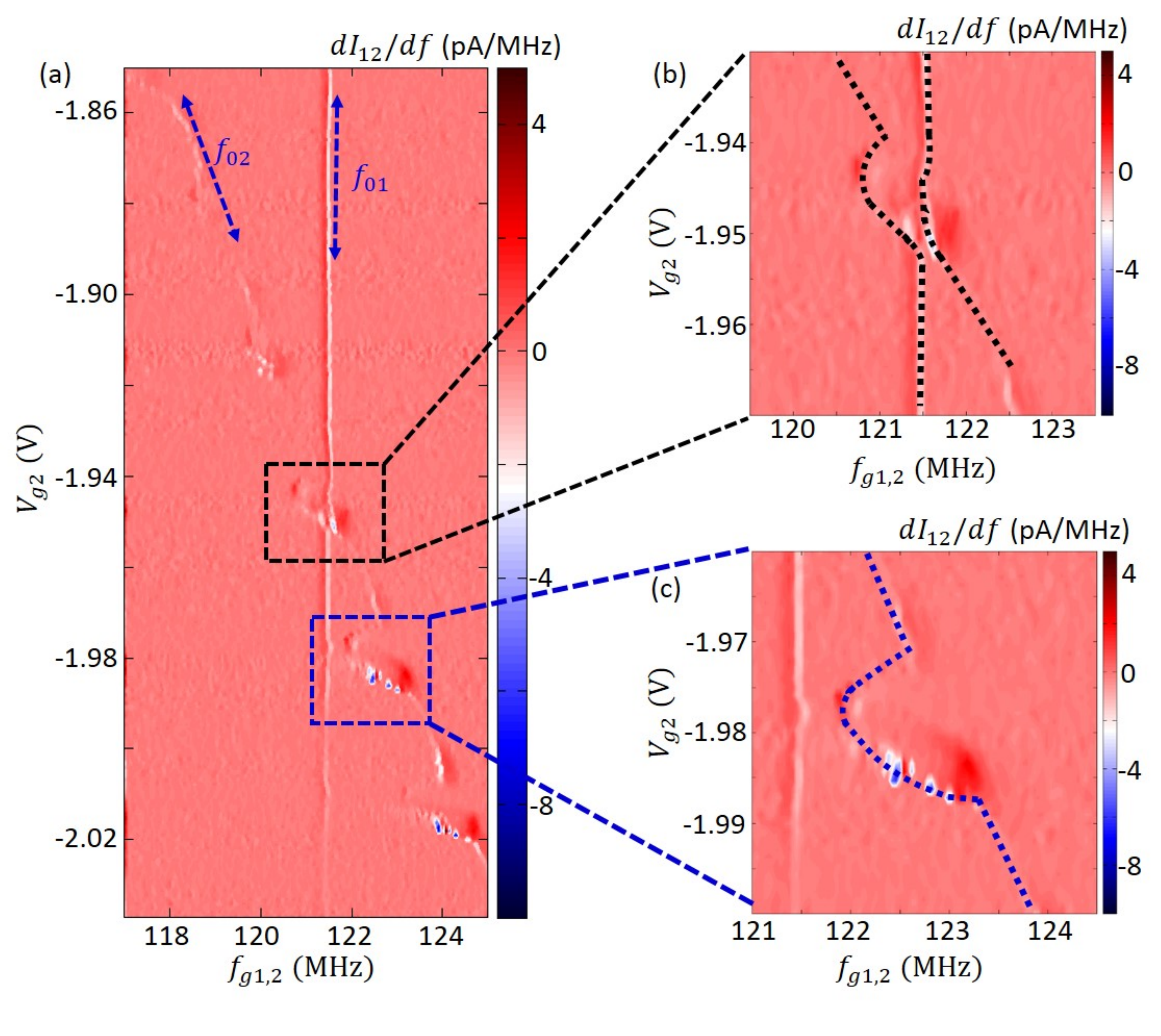}} \protect\protect\protect\caption{\textbf{Strong electron-phonon coupling.} (a) Color map of the coupled
system as a function of driving frequency and $V_{g2}^{{\rm DC}}$.
$V_{g1}^{{\rm DC}}$ is fixed at $-$0.657 V, and the current $I_{12}$
is measured between the D1 and D2 gates. The vertical line represents
resonator 1, while the transverse lines correspond to the electron
tunneling peaks of resonator 2. The oblique line with a negative slope
represents the resonance of resonator 2. (b) Zoom-in of the crossing
region for the two resonators and the current peak. (c) Zoom-in of
the interaction between resonator 2 and its electron tunneling. }
\end{figure}

Based on the fact that the resonators are highly tunable, we can study
the electron-phonon strong coupling by varying the gate field \cite{Steele2009,Lassagne2009,Meerwaldt2012,Moser2013,Benyamini2014}.
Fixing $V_{g1}^{{\rm DC}}=-0.657$ V, and the corresponding resonance
frequency $f_{01}=\omega_{m,1}/2\pi=121.7$ MHz, we scan $V_{g2}^{DC}$
to make $f_{02}$ near-resonant with $f_{01}$ and record the current
$I_{12}$. In this case, we drive the two resonators simultaneously
with two individual microwave sources at the same frequencies $f_{g1,2}$
($-$43 dBm for resonator 1 and $-$49 dBm for resonator 2). To achieve
a better resolution, we show the numerically-differentiated $dI_{12}/df$
as a function of frequency $f_{g1,2}$ and $V_{g2}^{{\rm DC}}$ in
Fig.~3(a). As indicated by the inclined arrow, $f_{02}$ linearly
decreases with increasing $V_{g2}$, and greatly modifies the current
for the bias field that yields the peaks or dips observed in Fig.~2(c).
At these points, we also observed a change of mechanical resonance
frequency of $f_{0,2}$. Such phenomena arise from the strong phonon-electron
tunneling interaction, which was firstly reported in 2009 \cite{Steele2009,Lassagne2009}.
The fluctuation of electron charges on the CNT induces the back-action
force on the mechanical modes, softening and damping the phonon modes
{[}Fig. 3(c){]}. The largest frequency shift is about $0.8\,\mathrm{MHz}$,
and the quality factors of the resonators are also largely reduced
from 10,000 to 500 because of damping, corresponding to an increase
of linewidth of phonon mode to $240\,\mathrm{kHz}$. The frequency
shift is about 3 times of magnitude larger than the linewidth of phonon
mode, verifying the strong coupling of the mechanical motion and single-electron
tunneling, and the damping rate induced by electron-phonon coupling
$\gamma_{e-{\rm ph}}/2\pi\sim240$ kHz, showing that the mechanical
motion is largely damped at these points.

In contrast to previous results, obtained for single-nanotube mechanical
oscillators, our system shows an additional vertical line (corresponding
to resonator 1) where the frequency does not change with $V_{g2}^{{\rm DC}}$.
This line clearly demonstrates the influence of resonator 2 on the
electron charge in resonator 1, which provides evidence of the non-local
control of the electron charge by phonons. When this line encounters
the photon-electron tunneling interaction frequency, as shown in Fig.~3(b),
the spectrum exhibits distinct features. A magnified 2D spectrum is
shown in Fig.~4(a), with a clear avoided-crossing when scanning the
frequency of resonator $2$ by varying $V_{g2}^{{\rm DC}}$. 
\begin{figure}[ptbh]
\centerline{\includegraphics[width=1\columnwidth]{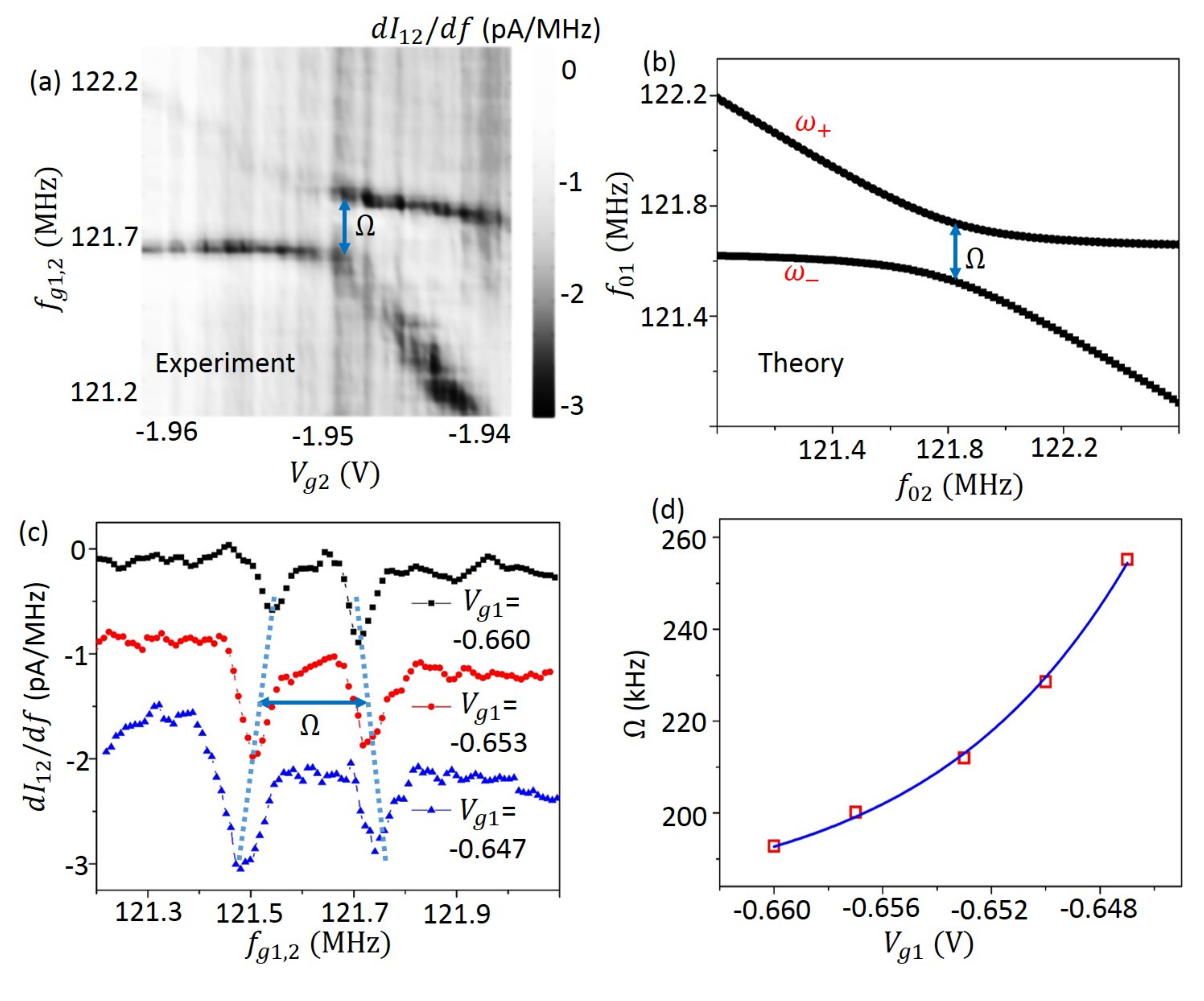}} \protect\protect\protect\caption{\textbf{ Strongly-coupled mechanical resonators} (a) Color map of
the coupled resonators, where a clear avoided crossing of two resonance
branches when the frequency of resonator 2 approaches the resonance
frequency of resonator 1. The splitting frequency, $\Omega$, represents
the coupling strength of the system. (b) Theoretical calculation of
the hybrid mechanical mode frequencies of the coupled resonators.
(c) The spectrum of the current derivative as a function of the driving
frequency for various gate voltages. (d) Coupling strength as a function
of the gate voltage $V_{g1}^{{\rm DC}}$.}
\end{figure}

To quantitatively verify this phonon-phonon interaction mechanism,
we also theoretically modeled the system using the Hamiltonian ($\hbar=1$)
\begin{equation}
\mathcal{H}_{m}=\omega_{m,1}a_{1}^{\dagger}a_{1}+\omega_{m,2}a_{2}^{\dagger}a_{2}+\frac{\Omega}{2}(a_{1}^{\dagger}a_{2}+a_{1}a_{2}^{\dagger}),
\end{equation}
where $\Omega$ is the phonon hopping between two resonators arising
from the tunneling. Therefore, the coupling induced new normal modes
with hybridization of two oscillators 
\begin{equation}
A_{\pm}=\frac{1}{\mathcal{N}}[(\omega_{\pm}-\omega_{m,2})a_{1}+\frac{\Omega}{2}a_{2}]\label{eq:hybridmodes}
\end{equation}
with normalization factor $\mathcal{N}=\sqrt{(\omega_{\pm}-\omega_{m,2})^{2}+\frac{\Omega^{2}}{4}}$
and eigenfrequencies 
\begin{equation}
\omega_{\pm}=\frac{1}{2}(\omega_{m,1}+\omega_{m,2}\pm\sqrt{(\omega_{m,1}-\omega_{m,2})^{2}+\Omega^{2}}).\label{eq:splitting}
\end{equation}
The calculated hybrid modes are shown in Fig. 4(b), by fitting the
parameters from our system. The good correspondence with experiment
confirms that the mechanism observed in Fig.~4(a) is coherent mechanical
mode coupling. Especially, the obtained $\Omega/2\pi\approx200\thinspace\mathrm{kHz}$
is an order of magnitude larger than $\kappa_{1,2}/2\pi\approx10\thinspace\mathrm{kHz}$.
Therefore, the phonon-phonon coupling is in the strong-coupling regime
$\Omega\gg\kappa_{1,2}$, and can be used for further coherent manipulation
\cite{Faust2013}.

The coherent coupling is further studied for various $V_{g1}$. As
indicated by Eq.~\ref{eq:splitting}, the coupling strength can be
extracted directly through the minimum frequency difference between
the hybrid modes ($\omega_{m,1}-\omega_{m,2}=0$). Therefore, we plotted
the spectrum of the minimum frequency splitting for different biased
$V_{g1}$. Note that the $V_{g1}$ changes the intrinsic frequency
of $\omega_{m,1}$, thus the $V_{g2}$ is adjusted to match the $\omega_{m,2}$
for each plot. Again, the two symmetrical dips in the spectrum confirm
the equal superposition of $a_{1}$ and $a_{2}$ (Eq. \ref{eq:hybridmodes})
for $\omega_{m,1}-\omega_{m,2}=0$. Intriguingly, the coupling strength
$\Omega$ shows a dependence on $V_{g1}$. This might be attributed
to the increase of tension of the nanotubes with increasing $V_{g1}$,
which reduces the evanescent phonon field on the S-electrode and suppresses
the phonon tunneling. We find that the coupling strength $\Omega$
can be tuned from 190 kHz to 250 kHz when $V_{g1}$ is varied from
$-0.66$ V to $-0.647$ V (Fig. 4(d)).

\subsection{Phononic quantum bus and discussion}

\begin{figure}[tbph]
\includegraphics[width=1\columnwidth]{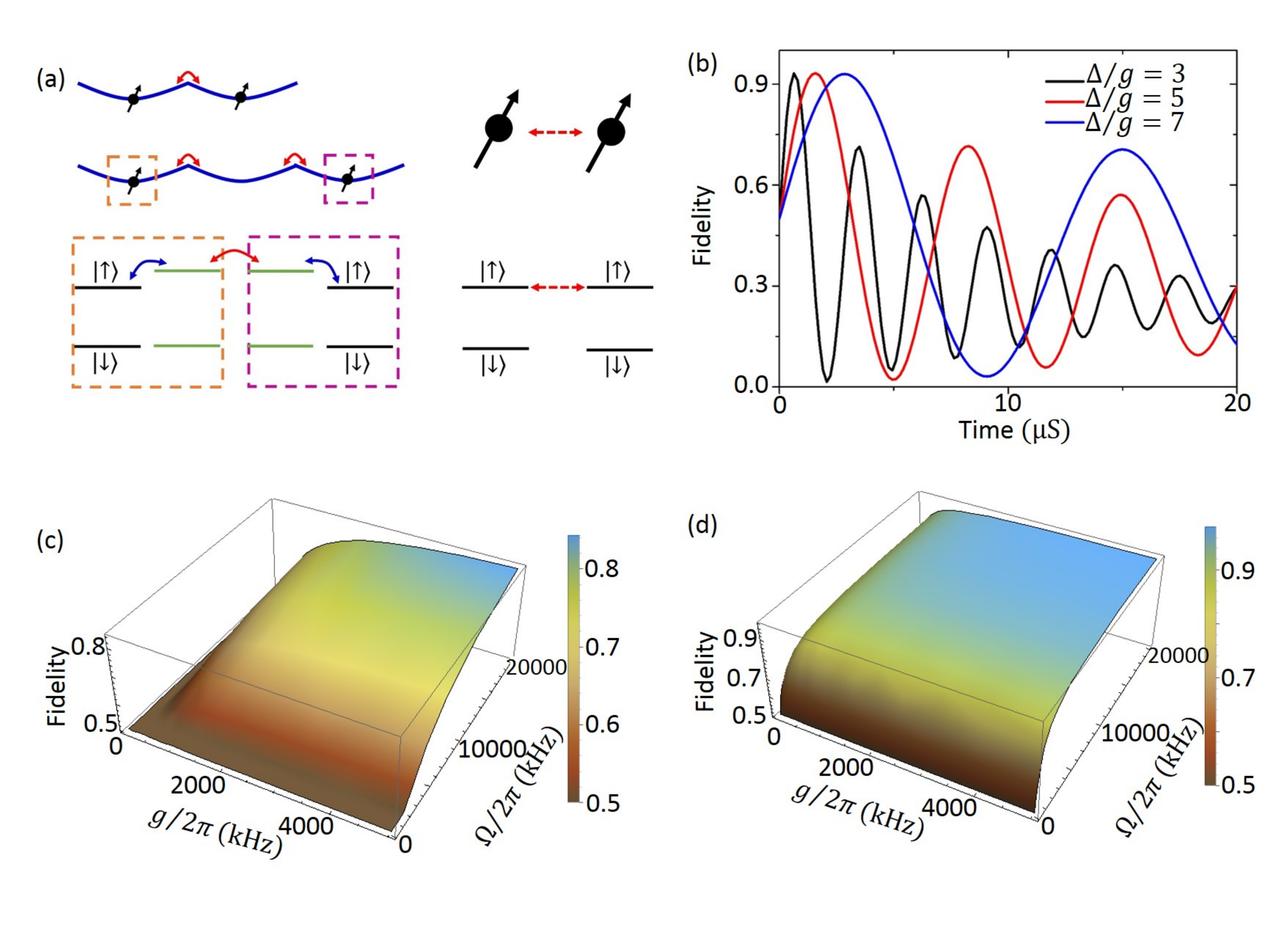} \protect\protect\protect\protect\caption{\textbf{ Phonon bus for spin qubits.} (a) Schematic diagram of the
effective spin-spin interaction mediated by phonons, whose frequencies
are largely detuned from the spin transition frequency. (b) Fidelity
of the remote entangled spin state $\left|\psi\right\rangle =\left|\downarrow\uparrow\right\rangle -i\left|\uparrow\downarrow\right\rangle $
generation for different detunings, with $g/2\pi=3000\thinspace\mathrm{kHz}$,
temperature $T=30\thinspace\mathrm{mK}$ and $\Omega=5000\thinspace\mathrm{kHz}$,
and other parameters of mechanical resonators from our experiments.
(c, d) show the fidelity versus spin-phonon coupling strength $g$
and $\Omega$, for $T=300$ mK (c) and $30$ mK (d).}
\end{figure}

Taking advantage of our novel fabrication method, the coupled CNT
mechanical resonators could be extended to one dimensional resonator
arrays, where a single CNT is put on a chip with multiple suspended
sections. Combined with an electrically-configurable individual mode
frequency and hopping rate, phonons could be manipulated and guided
along the array on demand. Our study suggests that phonons are promising
for quantum bus that can be used to transfer information over distances.
It is instructive to consider two possible configurations: Given double
quantum dots on each suspended section, phonons can assist single-electron
tunneling between the dots, as demonstrated in Fig.~3. As a result,
the quantum charge state can be coherently coupled with the localized
phonon. Alternatively, according to the inherent curvature and the
spin-orbit interaction \cite{Bulaev2008,Rudner2010,Steele2013}, the
spin states can be flipped by the phonons. Based on the phonon-phonon
coupling demonstrated in this work, the excitation can be transferred
between qubits over distance, and an effective qubit-qubit coupling
can then be realized by the virtual collective phonon mode.

Specifically, we take the qubit as an example to illustrate the generation
of spin entanglement states through the phononic quantum bus. As schematically
shown in Fig. 5(a), there are two mechanical resonators each containing
a single spin. More generally, such effects also hold for an array
of resonators. For ideal conditions, without any dissipation or decoherence,
the initial state $\left|\downarrow\uparrow\right\rangle $ of the
system would evolve to the state $\left|\psi\right\rangle =\left|\downarrow\uparrow\right\rangle -i\left|\uparrow\downarrow\right\rangle $,
which is the maximum entangled state over a distance of the order
of microns. For practical parameters, spin-phonon coupling strength
$g/2\pi=3000\thinspace\mathrm{kHz}$ \cite{Ohm2012,Palyi2012} (also
see the supplementary materials), temperature $T=30\thinspace\mathrm{mK}$
and $\Omega=5000\thinspace\mathrm{kHz}$, the fidelity of the entanglement
state $\mathcal{F}=\mathrm{Tr}[\rho(t)\left|\psi\right\rangle \left\langle \psi\right|)$
for various detuning $\Delta$ is plotted in Fig.~5(b) (see Methods
for details). Owing to the trade-off between phonon relaxation $\gamma_{j}$
and effective phonon-mediated coupling $g_{{\rm eff}}$, a moderate
detuning $\Delta/g=7$ shows a longer coherence time and also maximum
fidelity. In Fig.~5(c, d), the fidelity versus $g$ and $\Omega$
is shown for $T=300$ mK and $30$ mK, respectively. The results indicate
that both strong phonon-phonon coupling and spin-phonon coupling are
important for remote-entanglement generation. Although the thermal-excitation
decoherence substantially degrades the entanglement fidelity, it is
still possible to observe this remote-entanglement effect in the current
experiment setup ($T=270$ mK) if we can increase the $\Omega$ to
$10000$ kHz by decreasing the width of $S$ {[}Fig.~1(a){]}. High-fidelity
remote-entanglement generation ($\mathcal{F}>0.95$) is promising
with $T=30$ mK, $g>500$ kHz and $\Omega>5000$ kHz, which are parameters
that can be readily realized with current technologies.

In summary, we have studied the electron-phonon coupling in strongly-coupled
carbon nanotube nanomechanical resonators. Our study suggests the
use of phonons as coherent quantum information carriers to mediate
the effective interaction between electron spins. The engineered phonons
can also be used to initialize, readout and manipulate the electron
spin states. Using phonons as flying qubits is the first step in the
exploration of multiple degrees-of-freedom electron-phonon systems.
Possible future works would investigate CNT mechanical resonator arrays
in phononic quantum memories \cite{Zhang2015}, also many-body interactions
\cite{Soykal2013} and the implementation of possible quantum error
correction schemes in coupled spin-phonon chains \cite{Waldherr2014}.
This system can even be further integrated with superconducting circuits
to construct a hybrid quantum machine \cite{Xiang2013,Deng2015a}.

\section*{Methods}

\textbf{Sample preparation }The CNTs were grown using the ethane CVD
method on a silicon substrate with trenches. The prepared CNTs were
all double- or triple-walled CNTs, with diameters ranging from 2 to
3 nm. After depositing TiO2 nanoparticles onto the suspended parts
of the CNTs for visualization, the inner shell of the CNTs were drawn
out and placed in their proper positions with high precision using
two homemade tips under an optical microscope. The electrodes and
alignment marks were fabricated on an undoped silicon chip with 500
nm oxide, by optical lithography followed by metal deposition (5 nm
Ti and 45 nm Au) with an electron-beam evaporator. Two gate electrodes
beneath the resonators were fabricated by electron beam lithography
(EBL) followed by metal deposition (5 nm Ti and 45 nm Au). Finally,
the contact electrodes (10 nm Ti and 190 nm Au) were fabricated, to
decrease the residual resists as much as possible. The EBL resists
used here were single-layered PMMA 950 A4 for the gates and double-layered
for the contacts. After the transfer process, electrical annealing
was used to improve the contact. The resistance of our devices was
typically several hundred $\mathrm{k}\Omega$ at room temperature.

\textbf{Detection} The driving AC signals were produced by two individual
analogue signal generators (Agilent E8257D), attenuated by 30 dB at
room temperature, then transmitted to the sample through Lake-Shore
cables. There was an approximately 5 dB attenuation in the cable;
however, we estimate a $\sim$3 dB error from sample to sample in
our setup. Gate and bias DC voltages are controlled by the DC ports
of a lock-in amplifier (SR830). AC and DC voltages are combined by
a bias-T (Anritsu K251). Current through the resonator was measured
by a multi-meter, after a pre-amplifier (SR570).

\textbf{Theory of phonon mediated entanglement} Considering the simple
two-resonator case studied here, we obtain the system Hamiltonian
as \cite{Ohm2012,Palyi2012,Wang2014} 
\begin{eqnarray}
\mathcal{H} & = & \mathcal{H}_{m}+\frac{1}{2}(\omega_{s,1}\sigma_{z,1}+\omega_{s,2}\sigma_{z,2})\nonumber \\
 &  & +g(\sigma_{1,-}a_{1}^{\dagger}+\sigma_{2,-}a_{2}^{\dagger}+\sigma_{1,+}a_{1}+\sigma_{2,+}a_{2}).
\end{eqnarray}
For simplicity, we assume the phonon modes and spins are identical
($\omega_{s,1}=\omega_{s,2}$ and $\omega_{m,1}=\omega_{m,2}$). In
the normal mode representation, we have $A_{\pm}=\frac{1}{\sqrt{2}}(a_{1}\pm a_{2})$,
whose frequencies differ by $\Omega$. Allowing the phonon modes to
be largely detuned from the spin transitions ($\omega_{s,1}=\omega_{m,1}-\Omega/2-\Delta$),
therefore mediates the spin-spin interaction but rarely absorb the
excitations. By adiabatic elimination of the phonon modes, the spins
obey the Master equation as $\frac{d}{dt}\rho=-i[\mathcal{H}_{eff},\rho]+\sum_{j=1,2}[\frac{\gamma_{j,ph}}{4}\mathcal{L}(\sigma_{j,z},\rho)+\frac{\gamma_{j}}{2}(n_{th}+1)\mathcal{L}(\sigma_{j,-},\rho)+\frac{\gamma_{j}}{2}n_{th}\mathcal{L}(\sigma_{j,+},\rho)]$,
where the effective Hamiltonian is 
\begin{equation}
\mathcal{H}_{eff}=g_{eff}(\sigma_{1,-}\sigma_{2,+}+\sigma_{1,+}\sigma_{2,-}),
\end{equation}
with $g_{eff}\approx\frac{g^{2}\Omega/2}{\Delta(\Delta+\Omega)}$
and Lindbald form $\mathcal{L}(o,\rho)=2o\rho o^{\dagger}-o^{\dagger}o\rho-\rho o^{\dagger}o$.
Here, the pure dephasing rate $\gamma_{j,ph}/2\pi=1\thinspace\mathrm{kHz}$
is estimated from hyperfine interaction \cite{Csiszar2014}, and the
intrinsic energy relaxation is neglected owing to the negligible environment
phonon density of state, and the effective energy relaxation rate
arising from $a_{1,2}$ is $\gamma_{j}\approx\frac{g^{2}\kappa(\Delta^{2}+\Delta\Omega+\Omega^{2}/2)}{\Delta^{2}(\Delta+\Omega)^{2}}$.

 \bibliographystyle{nature}
\bibliography{nanotube}

\begin{thebibliography}{10}

\bibitem{Saito1998}
Saito, R., Dresselhaus, G., and Dresselhaus, M.~S.
\newblock {\em Physical Properties of Carbon Nanotubes}.
\newblock World Scientific Publishing,  (1998).
\newblock ISBN 978-1-86094-093-4 (hb) ISBN 978-1-86094-223-5 (pb).

\bibitem{Ilani2010}
Ilani, S. and McEuen, P.~L.
\newblock {\em Annu. Rev. Condens. Matter Phys.}{ \bf 1}(1), 1--25 (2010).

\bibitem{Laird2015}
Laird, E.~A., Kuemmeth, F., Steele, G.~A., Grove-Rasmussen, K., Nygard, J.,
  Flensberg, K., and Kouwenhoven, L.~P.
\newblock {\em Reviews Of Modern Physics}{ \bf 87}(3), 703--764 (2015).

\bibitem{Biercuk2005}
Biercuk, M.~J., Garaj, S., Mason, N., Chow, J.~M., and Marcus, C.~M.
\newblock {\em Nano Lett.}{ \bf 5}(7), 1267--1271 (2005).

\bibitem{Sapmaz2006a}
Sapmaz, S., Meyer, C., Beliczynski, P., Jarillo-Herrero, P., and Kouwenhoven,
  L.~P.
\newblock {\em Nano Lett.}{ \bf 6}(7), 1350--1355 (2006).

\bibitem{Sapmaz2006}
Sapmaz, S., Jarillo-Herrero, P., Kouwenhoven, L.~P., and Zant, H. S. J. V.~D.
\newblock {\em Semicond. Sci. Technol.}{ \bf 21}(11), S52--S63 (2006).

\bibitem{Grove-Rasmussen2008}
Grove-Rasmussen, K., J\o~rgensen, H.~I., Hayashi, T., Lindelof, P.~E., and
  Fujisawa, T.
\newblock {\em Nano Lett.}{ \bf 8}(4), 1055--1060 (2008).

\bibitem{Jung2013}
Jung, M., Schindele, J., Nau, S., Weiss, M., Baumgartner, A., and
  Sch\"{o}nenberger, C.
\newblock {\em Nano Lett.}{ \bf 13}(9), 4522--4526 (2013).

\bibitem{Petersson2012}
Petersson, K.~D., McFaul, L.~W., Schroer, M.~D., Jung, M., Taylor, J.~M.,
  Houck, a.~a., and Petta, J.~R.
\newblock {\em Nature}{ \bf 490}(7420), 380--383 (2012).

\bibitem{Deng2015}
Deng, G.~W., Wei, D., Li, S.~X., Johansson, J.~R., Kong, W.~C., Li, H.~O., Cao,
  G., Xiao, M., Guo, G.~C., Nori, F., Jiang, H.~W., and Guo, G.~P.
\newblock {\em Nano Lett.}{ \bf 15}(10), 6620--6625 (2015).

\bibitem{Meunier2011}
Meunier, T. and Ba, C.
\newblock {\em Nature}{ \bf 477}(7365), 435--438 (2011).

\bibitem{Mcneil2011}
Mcneil, R. P.~G., Kataoka, M., Ford, C. J.~B., Barnes, C. H.~W., Anderson, D.,
  Jones, G. A.~C., Farrer, I., and Ritchie, D.~A.
\newblock {\em Nature}{ \bf 477}(7365), 439--442 (2011).

\bibitem{Laird2012}
Laird, E.~A., Pei, F., Tang, W., Steele, G.~A., and Kouwenhoven, L.~P.
\newblock {\em Nano Lett.}{ \bf 12}(1), 193--197 (2012).

\bibitem{Moser2014}
Moser, J., Eichler, A., Guttinger, J., Dykman, M.~I., and Bachtold, A.
\newblock {\em Nat. Nanotechnol.}{ \bf 9}(12), 1007--1011 (2014).

\bibitem{Steele2009}
Steele, G.~A., Huttel, A.~K., Witkamp, B., Poot, M., Meerwaldt, H.~B.,
  Kouwenhoven, L.~P., and Zant, H. S. J. v.~d.
\newblock {\em Science}{ \bf 325}(5944), 1103 (2009).

\bibitem{Lassagne2009}
Benjamin, L., Yury, T., Jari, K., David, G.-S., and Adrain, B.
\newblock {\em Science}{ \bf 325}, 1107 (2009).

\bibitem{Ando2000}
Ando, T.
\newblock {\em J. Phys. Soc. Japan}{ \bf 69}(6), 1757--1763 June  (2000).

\bibitem{Huertas2006}
Huertas-Hernando, D., Guinea, F., and Brataas, A.
\newblock {\em Phys. Rev. B}{ \bf 74}(15), 155426 (2006).

\bibitem{Kuemmeth2008}
Kuemmeth, F., Ilani, S., Ralph, D.~C., and McEuen, P.~L.
\newblock {\em Nature}{ \bf 452}(7186), 448--452 (2008).

\bibitem{Palyi2012}
Palyi, A., Struck, P.~R., Rudner, M., Flensberg, K., and Burkard, G.
\newblock {\em Phys. Rev. Lett.}{ \bf 108}(20), 206811 (2012).

\bibitem{Ohm2012}
Ohm, C., Stampfer, C., Splettstoesser, J., and Wegewijs, M.~R.
\newblock {\em Appl. Phys. Lett.}{ \bf 100}(14), 143103 (2012).

\bibitem{Wang2014}
Wang, H. and Burkard, G.
\newblock {\em Phys. Rev. B}{ \bf 90}(3), 035415 July  (2014).

\bibitem{Bulaev2008}
Bulaev, D.~V., Trauzettel, B., and Loss, D.
\newblock {\em Phys. Rev. B}{ \bf 77}(23), 235301 (2008).

\bibitem{Rudner2010}
Rudner, M.~S. and Rashba, E.~I.
\newblock {\em Phys. Rev. B}{ \bf 81}(12), 125426 (2010).

\bibitem{Benyamini2014}
Benyamini, a., Hamo, A., Kusminskiy, S.~V., von Oppen, F., and Ilani, S.
\newblock {\em Nat. Phys.}{ \bf 10}(2), 151--156 January  (2014).

\bibitem{Zippilli2009}
Zippilli, S., Morigi, G., and Bachtold, A.
\newblock {\em Phys. Rev. Lett.}{ \bf 102}(9), 096804 mar  (2009).

\bibitem{Meerwaldt2012}
Meerwaldt, H.~B., Labadze, G., Schneider, B.~H., Taspinar, A., Blanter, Y.~M.,
  van~der Zant, H. S.~J., and Steele, G.~A.
\newblock {\em Phys. Rev. B}{ \bf 86}(11), 115454 (2012).

\bibitem{Wang2015}
Wang, H. and Burkard, G.
\newblock {\em Phys. Rev. B}{ \bf 92}(19), 195432 nov  (2015).

\bibitem{Rabl2010}
Rabl, P., Kolkowitz, S.~J., Koppens, F. H.~L., Harris, J. G.~E., Zoller, P.,
  and Lukin, M.~D.
\newblock {\em Nat. Phys.}{ \bf 6}(8), 602--608 May  (2010).

\bibitem{Gustafsson2014}
Gustafsson, M.~V., Aref, T., Kockum, A.~F., Ekstrom, M.~K., Johansson, G., and
  Delsing, P.
\newblock {\em Science}{ \bf 346}(6206), 207--211 September  (2014).

\bibitem{Schuetz2015}
Schuetz, M. J.~a., Kessler, E.~M., Giedke, G., Vandersypen, L. M.~K., Lukin,
  M.~D., and Cirac, J.~I.
\newblock {\em Phys. Rev. X}{ \bf 5}(3), 031031 (2015).

\bibitem{Liang2001}
Liang, W., Bockrath, M., Bozovic, D., Hafner, J.~H., Tinkham, M., and Park, H.
\newblock {\em Nature}{ \bf 411}(6838), 665--669 (2001).

\bibitem{Grove-Rasmussen2007}
Grove-Rasmussen, K., J\o~rgensen, H.~I., and Lindelof, P.~E.
\newblock {\em Phys. E Low-Dimensional Syst. Nanostructures}{ \bf 40}(1),
  92--98 (2007).

\bibitem{Huttel2009}
Huttel, A.~K., Steele, G.~a., Witkamp, B., Poot, M., Kouwenhoven, L.~P., and
  {Van Der Zant}, H. S.~J.
\newblock {\em Nano Lett.}{ \bf 9}(7), 2547--2552 (2009).

\bibitem{Rugar1991}
Rugar, D. and Grutter, P.
\newblock {\em Phys. Rev. Lett.}{ \bf 67}(6), 699--702 (1991).

\bibitem{Karabalin2009}
Karabalin, R.~B., Cross, M.~C., and Roukes, M.~L.
\newblock {\em Phys. Rev. B}{ \bf 79}(16), 165309 apr  (2009).

\bibitem{Mahboob2008}
Mahboob, I. and Yamaguchi, H.
\newblock {\em Nature Nanotech.}{ \bf 3}(5), 275--279 (2008).

\bibitem{Unterreithmeier}
Unterreithmeier, Q.~P., Weig, E.~M., and Kotthaus, J.~P.
\newblock {\em Nature}{ \bf 458}(7241), 1001--1004 (2009).

\bibitem{Moser2013}
Moser, J., Guttinger, J., Eichler, A., Esplandiu, M.~J., Liu, D.~E., Dykman,
  M.~I., and Bachtold, A.
\newblock {\em Nature Nanotech.}{ \bf 8}, 493 (2013).

\bibitem{Faust2013}
Faust, T., Rieger, J., Seitner, M.~J., Kotthaus, J.~P., and Weig, E.~M.
\newblock {\em Nature Phys.}{ \bf 9}(8), 485--488 (2013).

\bibitem{Steele2013}
Steele, G.~A., Pei, F., Laird, E.~A., Jol, J.~M., Meerwaldt, H.~B., and
  Kouwenhoven, L.~P.
\newblock {\em Nat. Commun.}{ \bf 4}, 1573 (2013).

\bibitem{Zhang2015}
Zhang, X., Zou, C.-L., Zhu, N., Marquardt, F., Jiang, L., and Tang, H.~X.
\newblock {\em Nat. Commun.}{ \bf 6}, 8914 nov  (2015).

\bibitem{Soykal2013}
Soykal, O.~O. and Tahan, C.
\newblock {\em Phys. Rev. B}{ \bf 88}(13), 134511 October  (2013).

\bibitem{Waldherr2014}
Waldherr, G., Wang, Y., Zaiser, S., Jamali, M., Schulte-Herbr\"{u}ggen, T.,
  Abe, H., Ohshima, T., Isoya, J., Du, J.~F., Neumann, P., and Wrachtrup, J.
\newblock {\em Nature}{ \bf 506}(7487), 204--7 February  (2014).

\bibitem{Xiang2013}
Xiang, Z.-L., Ashhab, S., You, J., and Nori, F.
\newblock {\em Rev. Mod. Phys.}{ \bf 85}(2), 623--653 apr  (2013).

\bibitem{Deng2015a}
Deng, G.-W., Wei, D., Johansson, J.~R., Zhang, M.-L., Li, S.-X., Li, H.-O.,
  Cao, G., Xiao, M., Tu, T., Guo, G.-C., Jiang, H.-W., Nori, F., and Guo, G.-P.
\newblock {\em Phys. Rev. Lett.}{ \bf 115}(12), 126804 sep  (2015).

\bibitem{Csiszar2014}
Csiszar, G. and Palyi, A.
\newblock {\em Phys. Rev. B}{ \bf 90}(24), 245413 (2014).

\end{thebibliography}

\section*{Acknowledgment}

We thank Liang Jiang, Lin Tian and Guido Burkard for beneficial discussions.
This work was supported by the National Fundamental Research Program
(Grant No. 2011CBA00200), the Strategic Priority Research Program
of the Chinese Academy of Sciences (Grant No. XDB01030000), and the
National Natural Science Foundation (Grants No. 11222438, 11174267,
61306150, 11304301, and 91421303). This work was also supported by
the National Basic Research Program of China (2012CB932301), National
Key Basic Research Program of China (MOST 2013CB922003) and NSF of
China (No. 11474178).

\section*{Author Contribution}

G.W.D. and D.Z. conceived the device. D.Z., X.H.W., K.L.J., D.L. and
Y.L. fabricated the samples. G.W.D., H.O.L., G.C., G.C.G. and M.X.
performed the measurements. G.W.D., D.Z. and X.H.W. analyzed the data.
C.L.Z. and G.P.G. conducted the theoretical investigation. G.P.G.
supervised the project. All authors contributed to the writing of
this paper.

\section*{Additional information}

\textbf{Competing financial interest: }The authors declare that they
have no competing financial interests. 
\end{document}